\documentclass[aps,pre,twocolumn,amsmath,amssymb,a4paper]{revtex4-1}

\usepackage{graphicx}
\usepackage{mathtools}
\usepackage{color}
\usepackage{tabularx}
\usepackage{float}

\usepackage{siunitx}
\usepackage{gensymb}

\usepackage{tikz}
\usepackage[permil]{overpic}
\usepackage{svg}

\begin{document}
\title{Designing the self-assembly of arbitrary shapes using minimal complexity building blocks}

\author{Joakim Bohlin$^{a,b}$}
\author{Andrew J. Turberfield$^{a,c}$}
\author{Ard A. Louis$^{d}$}
\author{Petr Šulc$^{b,}$}
\email[E-mail: ]{psulc@asu.edu}

\affiliation{$^a$Clarendon Laboratory, Department of Physics, University of Oxford, Parks Road, Oxford OX1 3PU, UK \\
$^b$ School of Molecular Sciences and Center for Molecular Design and Biomimetics, The Biodesign Institute, Arizona State University, 1001 South McAllister Avenue, Tempe, AZ 85281, USA \\
$^c$ Kavli Centre for Nanoscience Discovery, University of Oxford, Dorothy Crowfoot Hodgkin Building, South Parks Road, Oxford OX1 3QU, UK \\
$^d$ Rudolf Peierls Centre for Theoretical Physics, Department of Physics, University of Oxford, Keble Road, Oxford OX1 3NP, UK }

\begin{abstract}
The design space for a self-assembled multicomponent objects ranges from a solution in which every building block is unique to one with the minimum number of distinct building blocks that unambiguously define the target structure. Using a novel pipeline, we explore the design spaces for a set of structures of various sizes and complexities. To understand the implications of the different solutions, we analyse their assembly dynamics using patchy particle simulations and study the influence of the number of distinct building blocks and the angular and spatial tolerances on their interactions on the kinetics and yield of the target assembly. We show that the resource-saving solution with minimum number of distinct blocks can often assemble just as well (or faster) than designs where each building block is unique. 
We further use our methods to design multifarious structures, where building blocks are shared between different target structures.
Finally, we use coarse-grained DNA simulations to investigate the realisation of multicomponent shapes using DNA nanostructures as building blocks.


\end{abstract}


\maketitle




One of the hallmarks of all living systems is the self-assembly of functional multi-component nano- and micro-scale structures of well-defined sizes and shapes from molecular building blocks such as proteins, lipids, and RNA.  The field of bionanotechnology has long looked towards biology as an inspiration for the realisation of complex nanostructures.
There is very active interdisciplinary research into the principles and practice of the design of multicomponent self-assembled nanostructures from various building blocks that include proteins \cite{zhu2021protein}, DNA-coated nanoparticles \cite{liu2016self,xiong2020three,wang2012colloids}, as well as DNA nanostructures. 
Experimentally realised multi-component structures include unbounded large-scale assemblies with repeating periodic motifs such as 2D self-assembling tiles \cite{rothemund2004algorithmic}, 3D DNA crystals and lattices \cite{simmons2016construction,tian2020ordered}, as well as finite-sized assemblies such as icoshedral capsids \cite{sigl2021programmable}. 
Given the ability to functionalise DNA (e.g.\ by attaching proteins, gold nanoparticles, metallisation, or mineralisation \cite{nguyen2020ultrathin,hurst2006maximizing,chen2018dna}), controlled self-assembly of micrometer scale objects from nanoscale addressable blocks would have profound impact on our ability to realise structures with applications that include meta-materials for optical sensing and computing, 3D electronic circuit manufacture, and molecular factory construction \cite{hui2022dna,molesky2018inverse,heuer2021engineering}. Since DNA origami nanostructures \cite{rothemund2006folding}, the most popular building blocks in DNA nanotechnology, are limited in size by the scaffold oligonucleotide on which they are built, larger structures are achieved by joining multiple origami-based blocks, either by hybridisation of complementary stands \cite{tikhomirov2017fractal, nano-chambers_lin2020, tian_octahedra2020}, or by shape complementarity \cite{wagenbauer2017gigadalton, sigl2021programmable}. Recent advances have shown the assembly of finite-size 2D patterns from multiple DNA origamis \cite{tikhomirov2017fractal,wintersinger2022multi} reaching up to micrometre scale. Each position in the final assembly is occupied by a distinct DNA origami component, each of which must be prepared separately. Thus, a design with a minimal amount of distinct components would be beneficial.

Inverse design problems, where one tries to find a set of components that self-assemble to form a target structure, are often difficult to solve. Previous work has addressed the inverse  problem of designing isotropic interactions between spheres that lead to a desired target shape \cite{zeravcic2014size,arkus2009minimal,halverson2013dna}. Patchy-particle models have been used to study non-isotropic interactions, limited to specific particle regions (\textit{patches}) which only bind if they are within tolerated ranges of distance and angle \cite{zhang2004self}. Such models have been popular in studies of a wide range of soft matter systems and phase transitions, including water as well as colloidal systems forming periodic lattices \cite{romano2010phase,russo2021physics}. Finite-size systems studied include self-assembled particle clusters \cite{wilber2007reversible}, viral capsids \cite{hagan2014modeling,johnston2010modelling}, and other multiprotein assemblies \cite{villar2009self}. 

\begin{figure}
    \centering
    \begin{overpic}[width=\linewidth]{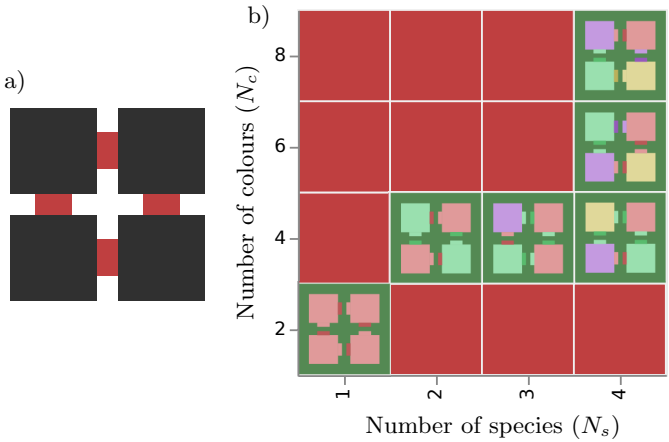}
        \put(-10,480){a)}
        \put(360,580){b)}

        \put(360, 320){\makebox(0,0){\rotatebox{90}{Number of colours (\(N_c\))}}}
        \put(540,-50){Number of species (\(N_s\))}
    \end{overpic}
    \vspace{0.5em}
    \caption{Assembly of a two-dimensional \(2 \times 2\) square polyomino using tile sets with different degrees of complexity. \textbf{a)} The input shape, consisting of four connected tiles. Each 2D tile can have up to four patches (one on each side): here they each have only two. \textbf{b)} Solution landscape. The green region shows possible assembly solutions, from the \emph{minimal set solution} using a single species and two colours (bottom left), to the \emph{fully addressable solution} using four species and eight colours (top right). The number of colours increases in pairs because patches are complementary. There are no possible solutions for the combination of \(N_s\) and \(N_c\) that are coloured in red (empty).}
    \label{fig:addressable}
\end{figure}

Here we show that a range of possible assembly sets are usually available for a given modular structure. The upper bound is the \textit{fully addressable} (maximum complexity) set, where every species and every pairwise interaction
is unique. This approach is, for example, used in single-stranded DNA tile designs \cite{ke2012three}. The other extreme is the \textit{minimal set}, using as few species and interaction types as possible. The latter is appealing for reasons of elegance and economy and because it may be more robust, for example, to cross-talk between nominally orthogonal interfaces \cite{huntley2016information}.  Our method can also be used to measure a structure's "designability", defined as the minimum number of blocks required to self-assemble it in high yield \cite{johnston2011evolutionary}. This quantity is related to Kolmogorov complexity \cite{soloveichik2007complexity,ahnert2010self}. It was recently proposed that natural systems (e.g.\ multiprotein complexes) are intrinsically biased towards minimal solutions, because these need fewer mutations to create and are hence easier to discover by random evolutionary search \cite{johnston2021}. Figure~\ref{fig:addressable} illustrates the maximum and minimum set concept by showing possible solutions for the assembly of a two-dimensional \(2 \times 2\) polyomino square. Two patches can only interact if they have complementary colours and orientations: the parameters of interest are the numbers of distinct tile species (\(N_s\)) and patch colours (\(N_c\)). For nanotechnology applications, specific patch interactions can be implemented e.g.\ through DNA single-stranded overhang design, where only complementary strands can bind, or by shape-complementarity, where only pieces with matching structures can connect. In this example the fully addressable solution uses four species and four connecting colour pairs (eight colours), as seen in the top-right of Figure~\ref{fig:addressable}.b. However, other solutions are also possible, with the simplest using a single species as seen in the bottom-left of the solution landscape. Other regions of parameter space, for example, using one species and four colours, are not topologically accessible.

We introduce a systematic method for solving inverse design for 3D finite-sized objects based on a SAT-assembly algorithm (mapping of inverse design to a Boolean Satisfiability Problem) \cite{russo2021physics,romano2020designing,rovigatti2022simple}.  We start with a simplified lattice model (a generalisation of the 2D polyomino model \cite{johnston2011evolutionary,ahnert2010self}) to explore the design space,  then investigate candidate solutions using simulations of the patchy particle model. We find that fully addressable designs work less well for patches with rigorous alignment criteria, i.e.\ where structure need to be precisely aligned for patches to bind. 
Next, we show that our inverse design pipeline can be used to design multifarious structures \cite{murugan2015multifarious,johnston2011evolutionary,sartori2020lessons}, where building blocks are shared between different target designs. 
Finally, we provide an interactive design pipeline that can translate any of our abstract patchy particle designs into DNA nanotechnology constructs that are amenable to experimental realisation.




\begin{figure*}[ht]
    \centering
    \begin{Overpic}[]{
        \resizebox{17.8cm}{!}{
            \input{figures/solve_pipeline}
            \includegraphics[width=0.65\linewidth]{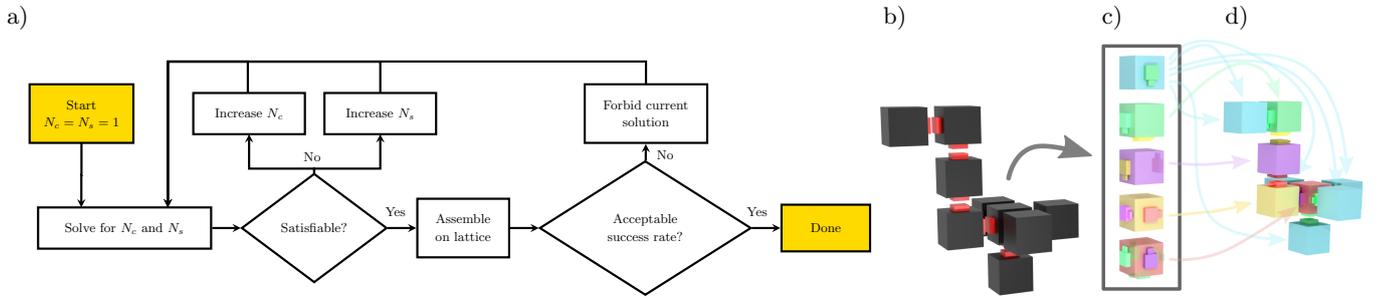}
        }
    }
        \put(0,200){a)}
        \put(640,200){b)}
        \put(800,200){c)}
        \put(890,200){d)}
    \end{Overpic}
    \caption{Polycube design. \textbf{a)} Algorithm for finding the minimal solution using SAT. Even if a solution is found to be satisfiable it might not assemble correctly every time. Additional solutions for a given \(N_c\) and \(N_s\) can be found by explicitly forbidding the current solution. \textbf{b)} Example of a swan shape, determined to have the minimal solution shown in \textbf{c)}. The five species assemble as seen in \textbf{d)} when cubes with compatible patches are added stochastically to an initial seed.
    }
    \label{fig:sat_alg}
\end{figure*}

\section*{Results}
We first describe our pipeline for efficient scanning of the landscape of solutions to the inverse design problem. This provides a library of candidate designs that self-assemble into a given target structure.

\subsection*{Design pipeline}
Figure~\ref{fig:sat_alg}.a shows a flow chart of the design pipeline for producing the solution landscape, including the minimal valid solution for a target multi-component structure. Each combination of parameters, i.e.\ the numbers of distinct colours (\(N_c\)) and of distinct particle (building block) species (\(N_s\)), is treated separately.  In our formulation of the problem we specify that each colour has only one compatible colour, matching typical sequence pairing rules in DNA nanotechnology or domain-specific protein-protein interactions. We formulate the inverse design problem as a Boolean Satisfiability Problem (SAT), i.e.\ in terms of a set of binary variables that are subject to constraints formulated as AND and OR logic clauses \cite{russo2021sat,romano2020designing}. The logic clauses specify that the designed building blocks can be properly placed onto a 3D lattice in such a way that required bonds are formed between patches with correct orientations and compatible colours (see Methods for details). Formulating the design problem as a SAT problem allows us to use highly optimised SAT solver algorithms \cite{sorensson2005minisat,audemard2009glucose}, which either show that the problem is not solvable for a given \(N_s\) and \(N_c\), or provide a solution specifying the  colour and orientation assigned to each patch on each particle species and the interaction matrix of compatible colours.

For the designs studied here, the SAT solver typically takes between a few seconds to tens of minutes to find a solution (if it exists) on a 2.4 GHz CPU. For problems that have no solution, it might take tens of minutes to hours for the solver to prove that no solution exists. Since each \(N_s\) and \(N_c\) combination is solved independently, they can be run in parallel using multiple CPU cores. Larger shapes require hardware with adequate memory. The \(3 \times 3 \times 3\) solid cube minimal solution was found in 19 seconds, but for many higher \(N_s, N_c\) the allocated 16 GB of RAM was not enough, as can be seen in Supp.~Figure~S4.b. For large values of $N_s$, close to the fully addressable solution, we have also implemented an alternative to the SAT-based approach, substitution solving (see Methods), which empirically reduces the number of species by replacing two different species with the same patch configuration by one (similar to greedy algorithm approaches previously employed in design of 2D DNA tile abstract assembly models \cite{ma2008synthesis}): the reduced system is then tested for correct assembly.

The solution obtained from the SAT solver guarantees that the target structure is an internal energy minimum as it includes all particles with all bonds satisfied. However, there might exist kinetic traps, that impede the formation of the target structure, or alternative assemblies with lower free energies. We use stochastic assembly simulations to identify such states.  If competing assemblies are found to prevent a target structure from assembling in high yield, we modify the set of clauses to exclude them and iterate. 

Molecular dynamics simulations of patchy particle assembly (see Methods) can be time-consuming, taking from a few hours to several days (or longer) to reach equilibrium, depending on the system size and specific interaction parameters. To speed up the testing of solutions generated by the SAT solvers, we developed a stochastic model of polycube  assembly on a lattice (described in detail in Methods) to assess the assembly yield \cite{johnston2011evolutionary}. In each assembly step a particle is chosen at random and added to a compatible patch on the growing cluster. This process is repeated until no more particles can be added (there is no available patch to attach to) or until the size of the growing shape exceeds a cutoff threshold. The method does not allow the removal of particles that have already been attached and thus corresponds to a system at a low temperature where the bond formation is irreversible.

\begin{figure}[ht]
    \centering
    \begin{overpic}[width=\linewidth]{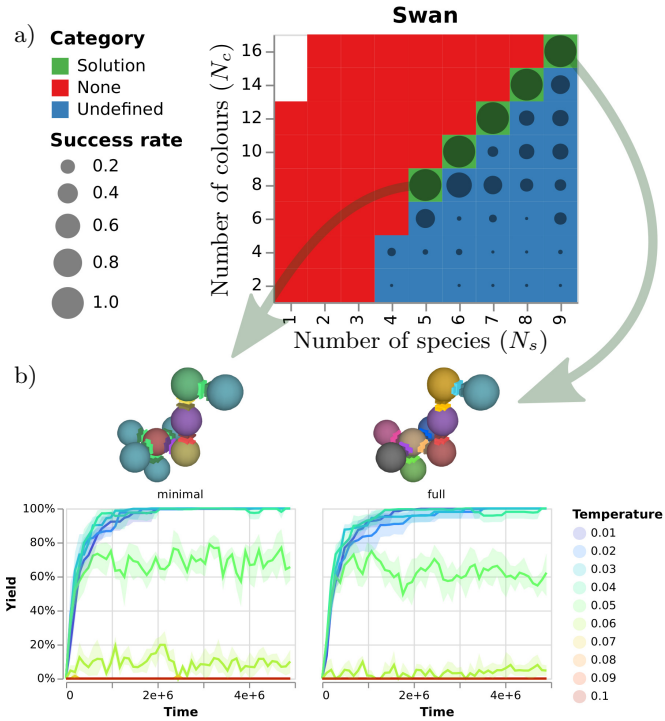}
        \put(10,950){a)}
        \put(10,470){b)}

        \put(300, 770){\makebox(0,0){\rotatebox{90}{Number of colours (\(N_c\))}}}
        \put(400, 520){Number of species (\(N_s\))}
    \end{overpic}
    \caption{Design and assembly of a polycube ``Swan''. \textbf{a)} Solution landscape for the ``Swan'' shape seen in Figure~\ref{fig:sat_alg}.b). Circle sizes indicate the assembly success rate found for the corresponding number of colours and species.\textbf{b)} Assembly kinetics for different ``Swan'' designs. The left column shows the minimal solution, while the right is the fully addressable solution. Solid lines are mean values from 5 duplicate simulations; error bands show the 95\% confidence interval. Each simulation uses the narrow type \(0\) potential (see Methods). Temperature and time are measured in simulation units.}
    \label{fig:swan}
\end{figure}

Figure~\ref{fig:sat_alg} illustrates the iterative application of stochastic assembly simulations in conjunction with a SAT solver to find acceptable solutions to an assembly problem.
This stochastic assembly simulation is repeated 100 times for each design, with each simulation taking only a fraction of a second of CPU time (for cutoff cluster size 100 particles). If the simulations do not consistently assemble the correct shape, we add Boolean clauses to exclude this solution from the SAT solver and generate a new one. If, after 100 trials, we still cannot find a valid solution, we assume that no valid solution can be found. In case some but not all of the 100 trials assembles correctly, we label the problem as {\it Undefined}, and record the fraction of stochastic  simulations leading to successful assembly (shown as success rate in Fig.~\ref{fig:swan}.a). If the SAT solver shows that no solution exists, we label the solution to the problem as {\it None}.

As an example of the design pipeline, Figure~\ref{fig:swan}.a shows the solution landscape for a ``swan'' shape. The abstract design on a 3D lattice is shown in Figure~\ref{fig:sat_alg}.b. While the fully addressable solution uses \(N_s=9\) species and \(N_c=16\) colours, the shape can also be reliably assembled with as few as \(N_s=5\) species and \(N_c=8\) colours.
Larger and more complex shapes require more time to rule out invalid solutions. See Supp.~Figures~S1-6 for more examples of shapes and their solution landscapes. 

\begin{figure*}[ht]
    \centering
    \begin{overpic}[width=17.8cm]{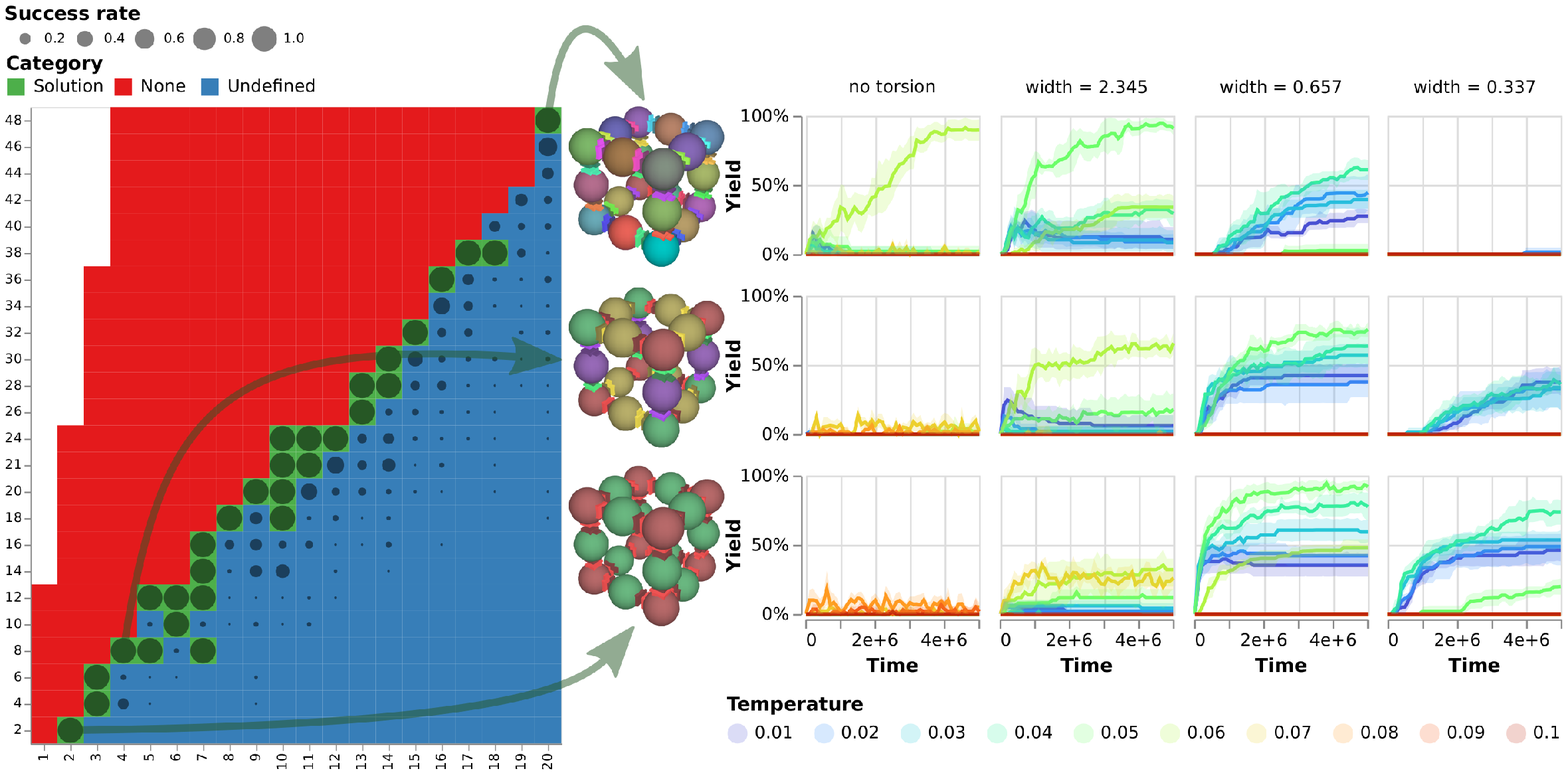}
        \put(-10,420){a)}
        \put(500,450){b)}

        \put(0, 240){\makebox(0,0){\rotatebox{90}{Number of colours (\(N_c\))}}}
        \put(110, -10){Number of species (\(N_s\))}
    \end{overpic}
    \caption{Design and assembly of a wireframe \(3 \times 3 \times 3\) cube. \textbf{a)} Solution landscape evaluated using a SAT solver and polycube lattice assembly simulations.  \textbf{b)} Assembly kinetics and yield explored using patchy-particle molecular dynamics simulations. The top row shows the fully addressable solution, using 20 species and 48 colours. The middle row shows a significantly smaller, but intermediate solution, using 4 species and 8 colours. The bottom row shows the minimal solution, using 2 species and 2 colours. Columns correspond to different interaction potentials: the leftmost column corresponds to wide patches with no torsional constraint. Columns to the right show torsional patches with decreasing patch widths. Solid lines are mean values from 5 duplicate simulations, and error bands show the 95\% confidence interval band. Each simulation has a \(0.1\) particle density.}
    \label{fig:cubePotentials}
\end{figure*}

\subsection*{Kinetics, yield and effects of torsional modulation}
\begin{figure}
    \centering
    \includegraphics[width=0.5\textwidth]{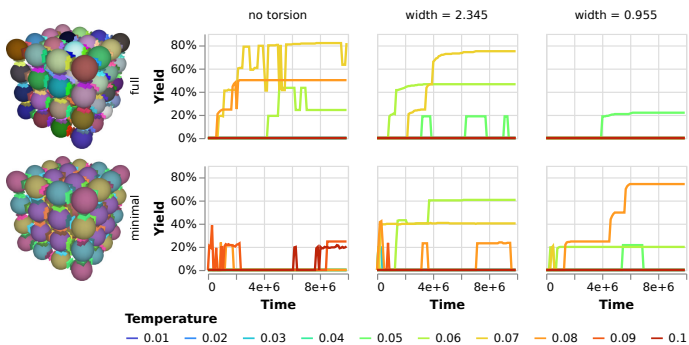}
    \caption{Assembly yield for two solid $4\times 4 \times 4$ cube designs. The top row shows the fully addressable solution, using 64 species and 288 colours. The bottom row shows the minimal solution, using 6 species and 16 colours. Columns correspond to different interaction potentials, with the leftmost column showing wide patches without torsional constraint. Columns to the right show torsional patches with decreasing patch widths. Each simulation has a $0.1$ particle density. The slightly different yield measure used, for reasons of efficiency, for the $4\times 4 \times 4$ cube underestimates some yields (see Methods).}
    \label{fig:cube4Potentials}
\end{figure}

Using our SAT-based pipeline to scan the solution space, we compare the assembly yields and kinetics for a set of designed target structures that include both symmetric and asymmetric shapes, of sizes up to 64 components (Figs.~\ref{fig:swan}, \ref{fig:cubePotentials}, \ref{fig:cube4Potentials} and Supplementary Figs. S7-9).

For each target shape we use a SAT solver and the stochastic polycube solver to evaluate the solution space.  Selected designs that assemble with 100\% success rate in the polycube solver, including both fully addressable and minimal solutions, are then tested using a molecular dynamics patchy particle simulations \cite{romano2020designing}. The patchy particle simulation model represents each particle as a sphere, with up to 6 patches on its surface arranged in a cubic geometry. Two patches can bind to each other if they have complementary colours and if their angular alignment lies within a specified interval (see Methods). All simulations start from a disassembled state consisting of unbound building blocks (in a box size corresponding to number density $0.1$) containing enough building blocks to assemble 5 to 10 copies of the target structure, depending on the target size. All building blocks are included in stochiometric ratio based on how often they are used in the target structure.

We use the molecular dynamics simulations to measure the kinetics and yield of the assembly and to identify possible misassembled states that were not captured by the polycube lattice assembly model (including undesired clusters or aggregates that do not fit onto a cubic lattice). 

For the swan design (Figure~\ref{fig:swan}.b), the building blocks of the minimal solution assemble just as well as the fully addressable assembly, using approximately half the number of species and colours. The same trend is seen in Supp.~Figure~S7, which shows corresponding results for ``robot'' and ``letter J'' shapes. The ``letter J'' shape's minimal solution uses only one species fewer than the fully addressable solution (Supp.~Figure~S1), so the differences are minor, but the minimal solution of the ``robot'' (Supp.~Figure~S2) requires fewer than half the number of species and colours of the fully addressable solution.
The designs in Fig.~\ref{fig:swan} and Supp.~Figs.~S1, S2, and S6 are all topologically self-limited, meaning that they stop growing because there are no more patches to bind. In graph theory terms, the topology of these structures (with building blocks as nodes and bonds as edges) are trees. Because of this, most of the possible competing alternative assemblies for topologically self-limited structures are successfully identified by the polycube assembly model.


However, spatially self-limiting structures (like the square in Figure~\ref{fig:addressable}) have loops in their topology. In such structures, misalignment due to flexibility in the bonds between particles can result in aggregation (see Supp.~Figure~S12 for an example). We can use the patchy particle simulations to investigate how a change in bond flexibility affects assembly yield and kinetics. We consider either patches where bound particles can freely rotate around the formed bond (``no torsion'' option), as well as a torsional potential where connection flexibility is modulated by the width of the angle-modulation potential (see Methods). 

To compare minimal, intermediate and fully addressable solutions, as well as the effects of bond flexibility, for designs which contain loops in their topology, we study the assembly of wireframe and filled cubes of side length 3 (Figures.~\ref{fig:cubePotentials} and S8-9) using different angular interaction potentials. For the wireframe cube, the fully addressable solution consists of 20 species; the minimal solution consists only of two. For filled cubes, the fully addressable solution requires 27 species, while the minimal has only 4. 

We note that narrower patch widths lead to slower assembly times for both filled and wireframe cube designs. Such an effect is expected, as encounters between particles with a narrow interaction potential are less likely to lead to a binding event. A wider patch is not necessarily better, however. As discussed above, increased flexibility allows off-lattice binding and decreases assembly yields. 

The tolerance of misalignment would depend on the physical realisation of patchy particles. Protein domains or shape-complementary DNA origami \cite{gerling2015dynamic} require interacting particles to be closely aligned, corresponding to very narrow potentials with small widths. On the other hand, hybridisation of single-stranded overhangs, commonly used to create multi-component assemblies with DNA nanotechnology, would typically allow more freedom to rotate around the formed bond.

Aggregation is also mitigated by cooperative binding, where particles form multiple bonds. This can be seen in Supp.~Figure~S9, where solid \(3 \times 3 \times 3\) cubes are assembled. The filled cube, in which particles have between three and six neighbours, offers higher binding cooperativity than the wireframe cube in which most particles have only two.
The structural constraints introduced by multiple interactions reduce misalignment and increase assembly yield for all solutions, including the minimal design, even in the absence of explicit torsional constraints.

For wider (or no torsion) patches, the fully addressable solution has the higher yield (at a particular optimal temperature for a given design) as it is less likely to form misassembled structures associated with patch flexibility. However, we observe that the minimal and intermediate solutions show faster assembly than the fully addressable solution as the interaction patches get narrower: fewer species makes it more likely that two nearby particles have compatible patches.

For a solid cube, the number of species required for a fully addressable solution scales as \(N_s = l^3\), where \(l\) is the dimension of the cube. With larger \(l\) the relative reduction in complexity of the minimal solution increases. While a fully addressable \(4\times4\times4\) cube needs 64 unique species, the minimal solution found with our design pipeline can assemble with only 6 species (Supp.~Figure~S5). Figure~\ref{fig:cube4Potentials} shows that the minimal solution can provide a high yield, at least provided that the patches are narrow enough. 
The stochastic assembler used in this work accepts shapes with all cubes in the designed positions, whether or not all possible connections are formed. The minimal solution for the \(4\times4\times4\) cube allows assemblies with all positions occupied but containing some mismatched bonds. The stochastic assembler can be programmed to disallow solutions producing mismatched bonds.
We verified, however, that even with mismatched bonds, the cubes are stable in molecular dynamics simulations. 

\subsection*{Multifarious assembly}

\begin{figure}
    \centering
    \includegraphics[width=\linewidth]{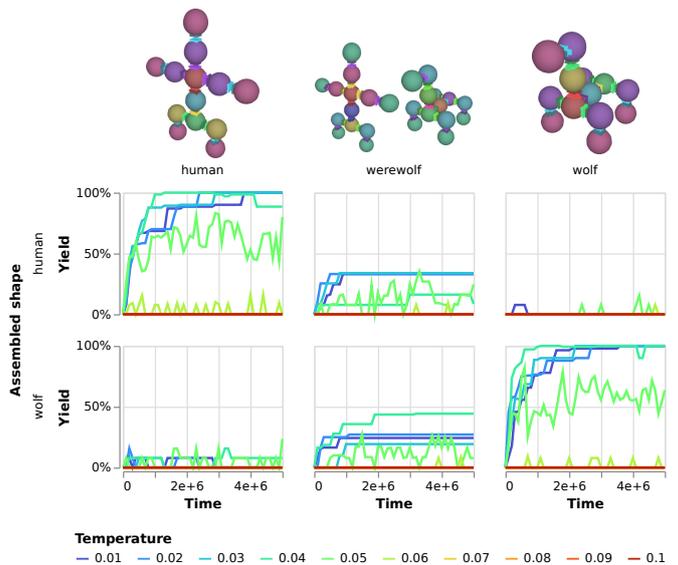}
    \caption{Multifarious assembly. The leftmost column shows the minimum solution for the ``human'' shape, the rightmost shows the minimum solution for the ``wolf'' shape, and the middle column shows the minimum solution that can reliably assemble both human and wolf shapes, here labelled ``werewolf''. The top and bottom rows show the yields of ``human'' and the ``wolf'' shapes respectively for the three sets of particles. Each simulation is done using the wide torsional patch potential with width = \(2.346\)) at a \(0.1\) particle density.}
    \label{fig:multifarious_sim}
\end{figure}

Another useful feature of the SAT solver approach is its ability to design \emph{multifarious} assemblies, that is, sets of building blocks that can assemble into more than one shape with at least some species shared between the possible products \cite{murugan2015multifarious, ahnert2010self}. To achieve this, we define the multiple distinct target shapes simultaneously in the input to the solver in our design pipeline. 

As an example, we use the pipeline to find a "werewolf" minimal solution that can assemble into two possible shapes: a "human" or a "wolf", consisting of 13 and 14 particles respectively (Figure \ref{fig:multifarious_sim}). Chimeric assemblies are treated in the same way as other off-target assemblies and are thus excluded by design. The multifarious assembly solution consists of ten species. Two species are shared between the shapes while each shape use an additional four species each. 
We test the obtained solution in patchy particle simulations. In Figure \ref{fig:multifarious_sim}, we compare the minimal assembly solutions generated for human and wolf shapes individually (which lead to 100\% assembly yields at sufficiently low temperatures) to the multifarious werewolf assembly solution. The relative concentration of species in  werewolf simulations was chosen so that there was sufficient material to assemble either 10 humans or 10 wolves. Both shapes form, 
with the wolf shape being slightly favoured at low assembly temperatures. 


\subsection*{Implementation with DNA nanotechnology}
\begin{figure}[h]
    \centering
    \includegraphics[width=\linewidth]{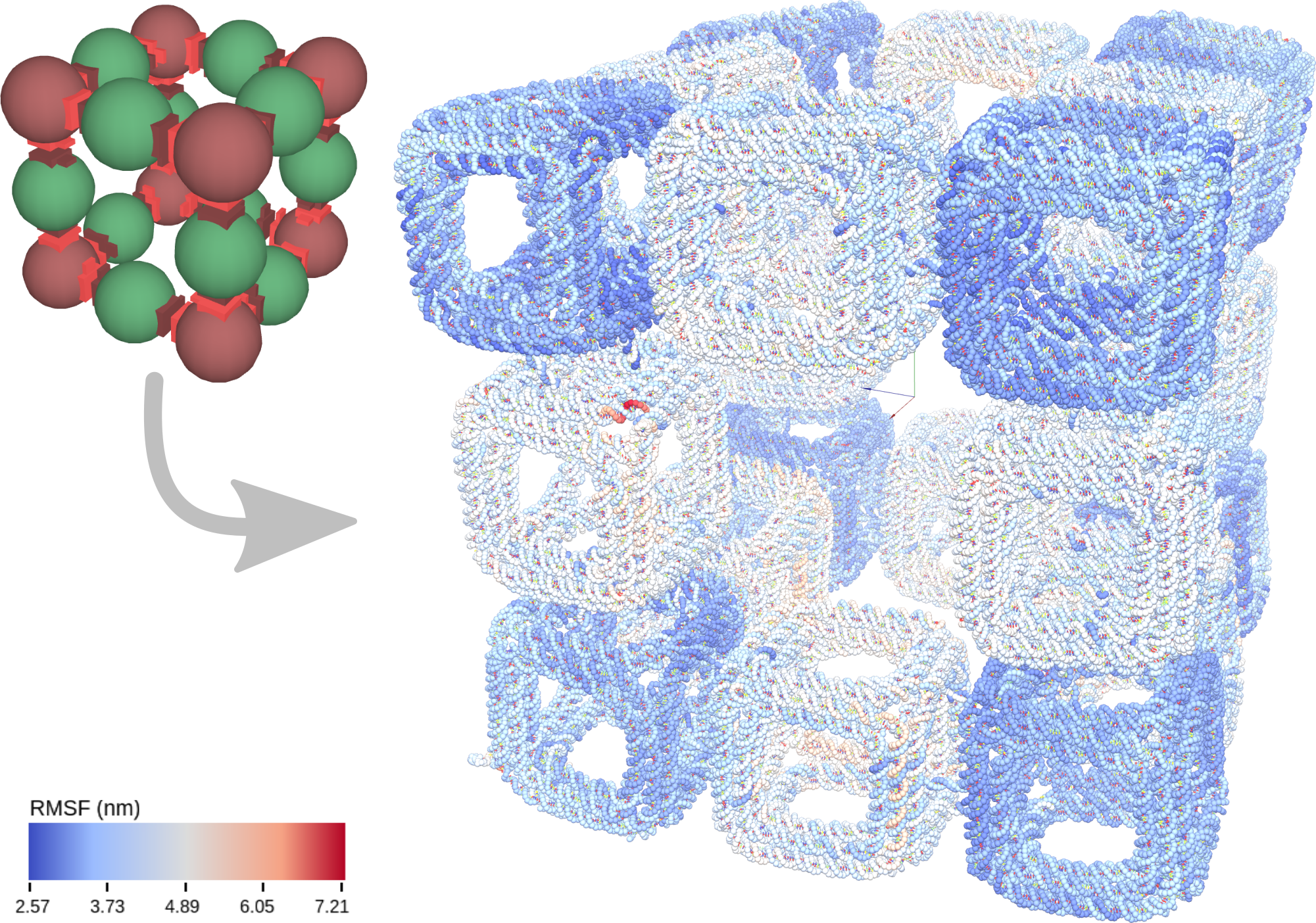}
    \caption{Mean structure and root-mean-square-fluctuation (RMSF) analysis (right) of the wireframe \(3 \times 3 \times 3\) cube minimal solution (top left), formed from DNA origami "patchy" particles and simulated using oxDNA for a duration corresponding to about $10^{-3}$ s.
    }
    \label{fig:hollowCubeDNA}
\end{figure}

To demonstrate a possible experimental realisation of the six-valent cubic-shaped designs considered here, we designed a wireframe DNA origami nanostructures using ATHENA \cite{jun2021rapid} and the oxView software tools \cite{poppleton2020design}. The design was verified with oxDNA \cite{vsulc2012sequence,Snodin2015,ouldridge2011structural}, a coarse-grained model of DNA. The DNA origami cubes can bind to each other through single-stranded (ssDNA) overhangs. One ``patch'' of our abstract polycube model corresponds to a set of ssDNA overhangs, each with a unique sequence, that are placed on the edges of a cube face, thus realising a patch with a well-defined orientation. Complementary patches are provided with complementary overhangs. Figure~\ref{fig:hollowCubeDNA} shows the structure, and structural fluctuations, of a minimal wireframe $3\times3\times3$ cube assembly realised using our origami design, as represented in an oxDNA simulation.

\section*{Discussion}

Our new automated design pipeline 
can explore the landscape of assembly solutions for arbitrary shapes. In many cases, the  minimum solution comprises significantly fewer particles than the target structure and can assemble more rapidly than the solution in which every component is unique, with comparable or higher yields.
Interestingly, the fact that, for highly specific binding domains such as those relevant for protein assembly, the minimal solution assembles better than more complex designs suggests an interesting evolutionary hypothesis:  The recently discovered non-adaptive bias towards minimal protein complex designs~\cite{johnston2021} may produce complexes that will assemble better, providing an additional adaptive evolutionary driver towards simpler and therefore more symmetric structures.

We found that for narrow (less flexible) patch interaction, the minimal solution assembles faster than the fully addressable solution. Hence, not only does the minimum assembly solution presents advantage in terms of manufacturing costs, for certain experimental realisations and conditions, it is also predicted to better assemble than the fully addressable option. Generally, the kinetics of assembly is slower for narrower patch interactions, as random collisions are less likely to lead to successful bond formation. Additionally, for the fully addressable system, two randomly chosen particles are less likely to have compatible interfaces, and hence we expect the structure growth to be slower than for the minimal solution.

Our results are likely to benefit the DNA nanotechnology field: the requirement for fewer unique components would mean significant savings in both time and resources, and we have developed a new automated design tool that can convert target structure design to a nucleotide-level coarse-grained DNA nanostructure representation for computational verification or for guiding an experimental design.  Our inverse design method is also applicable to other designed self-assembling systems, such as multi-component protein structures or coated nanoparticles \cite{zhu2021protein,xiong2020three}. We note that for experiment it may be desirable to incorporate specific building blocks, perhaps uniquely functionalised by a specific material coating or attachment of a guest molecule, at specific locations. Our design method can be extended to allow for inclusion of such constraints.


We note that here we focused on constant-temperature assembly. Constant temperatures are appropriate for potential \emph{in vivo} applications, but it is probable that even better assembly yields could be achieved using more complicated temperature protocols \cite{bupathy2022temperature}. We note that previous work has shown that very large multi-component systems, such as single-stranded  tiles with targets consisting of hundreds of different species, require controlled cooling from high temperature to ensure rapid assembly and high yields \cite{jacobs2015rational,fonseca2018multi}. The initial high temperature helps the system cross the nucleation barrier; subsequent cooling to low temperature is necessary to combat the high entropic cost of incorporating the last few missing particles into the growing structure. Our stochastic polycube model can be extended to allow for thermally activated building block detachment, thus enabling investigation of variable-temperature assembly. The design pipeline could also be adapted to investigate the design of different protocols, including staged assembly in multiple pots \cite{tikhomirov2017fractal}, the combination of different interaction strengths \cite{ma2019inverse}, or optimisation of order in which different species are gradually added. 


A further continuation would be to study how the complexity, in terms of the number of species and colours, affect the robustness to defects, such as missing interaction site. Minimal designs are typically more robust to random design errors~\cite{greenbury2016genetic, johnston2021}. In a minimal design, a faulty individual particle is expected to be more easily replaced by another particle of the same species. However, an error affecting an entire species is expected to have larger consequences for species used in multiple locations. 

Finally, the current design pipeline is hindered by the size of the search space. While solutions using few species and colours are readily discovered, larger and more complex shapes requiring large particle libraries need significantly more time and memory to find a solution. Future work could try to mitigate this by separating shapes into smaller sub-shapes, a strategy that could also be linked to staged assembly.
We will also explore in future work alternatives to the stochastic simulation to find possible competing alternative states, such as geometry optimisation within the computational potential energy landscapes \cite{wales1999global,fejer2018designing}.


The developed design tool, simulation software, as well as an interactive tool that converts abstract design into DNA nanostructure building blocks, are freely available at \href{https://github.com/Akodiat/polycubes}{https://github.com/Akodiat/polycubes}.

\section*{Methods}
\subsection*{The polycube model}
\label{sec:polycube_model}
Used as an initial test of potential solutions, the stochastic self-assembly polycube model can quickly assemble an input rule, discarding non-deterministic and unbounded rules as invalid.
The lattice-based model consist of cubes that bind together if the patches on their sides are compatible. 
Each patch has a colour of integer number \(n\) that will bind to \(-n\), and one of four possible orientations: \includegraphics[width=8pt]{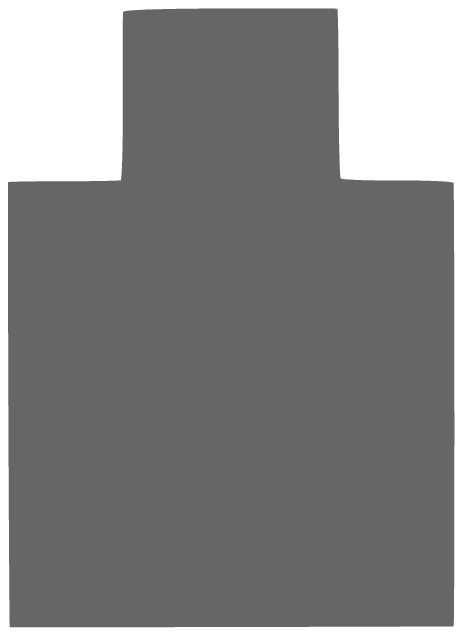}\hspace{4pt}(\(0\)), \begingroup\setbox0=\hbox{\includegraphics[width=8pt,angle=-90]{face.eps}}\parbox{\wd0}{\box0}\endgroup\hspace{4pt}(\(\frac{\pi}{2}\)), \begingroup\setbox0=\hbox{\includegraphics[width=8pt,angle=180]{face.eps}}\parbox{\wd0}{\box0}\endgroup\hspace{4pt}(\(\pi\)) or \begingroup\setbox0=\hbox{\includegraphics[width=8pt,angle=90]{face.eps}}\parbox{\wd0}{\box0}\endgroup\hspace{4pt}(\(\frac{3\pi}{2}\)). Each building block cube belongs to a species, with a set of species defined as the polycube rule. Supp.~Figure~S10 illustrates how a rule with two species is defined and assembled into a deterministic polycube structure.
The assemblies grow as new cubes with compatible species are stochastically added. First, the polycube stochastic assembly starts from a randomly chosen particle which is placed onto the lattice. Then next randomly chosen particle is attached to the previously placed particle such that their patches are compatible and form a bond. The algorithm always grows a single cluster and does not remove any particles that have already been placed, corresponding to a system at a low temperature where the bond formation is irreversible. It stops when no more particles can be added, or when the size of the cluster exceeds given threshold. This process is repeated 100 times for each tested rule. The coordinates of the cubes are compared to those of the desired shape, accounting for rotation and translation, and if the assembly process forms the same finite shape in all trials, then the rule is recorded as valid for assembling that shape.





\subsection*{Satisfiability solving}
\label{sec:SAT}


For any given polycube shape, it is trivial to find a fully addressable rule that assembles it. This is done by assigning a unique species to each particle, as well as a unique colour (and complementary colour) to each connection, in the shape. However, the more species you require, the more unique origami structures would need to be constructed, adding considerable effort to experimental applications. While a simpler polycube rule can be designed by hand, another option is to use a satisfiability solver.

In essence, it is possible to formulate a Boolean expression that, if true, means it is possible to assemble a given polycube topology using a given number of colours and species. Introducing the Boolean variables shown in Table 1, we formulate clauses to constrain the problem, as seen in Table 2. Clauses (i)-(vii) are the same as introduced in Refs.~\cite{romano2020designing,russo2021sat} while the remaining are added, together with variables \(x^D\), \(x^A\) and \(x^O\), to include torsional restrictions. The respective clauses listed in Table 2 ensure that the solution satisifies the following: (i) Each colour is compatible with \textit{exactly one} colour. (ii) Each patch has \textit{exactly one} colour. (iii) Each target shape position contains a single species with an assigned rotation. (iv) Adjacent patches in the lattice must have compatible colours. (v) Patches at a lattice position are coloured according to the (rotated) occupying species, where $\phi_r(p)$ corresponds to the patch number that will overlap with original patch number $p$, after one of 24 possible rotations $r$ (see Supp.~Table~S1) is applied that rotates the cubic-shaped particle in such a way that patches overlap. (vi) All \(N_t\) species are required to be present in the assembled target shape. (vii) All \(N_c\) patch colours are required to be used in the solution. (iix) Each patch is assigned \textit{exactly one} orientation. (ix) Adjacent patches in the target lattice must have the same orientation. (x) Patches at a lattice position are oriented according to the (rotated) occupying species. The SAT problem is conjunction of clauses (i)-(x). The clauses and variables are defined for all combinations of colours $c$, patches $p$, rotations $r$, species $s$, orientations $o$, positions $l$ in the target shape $l$. Clauses (iv) are only defined for adjacent positions in the target shape.

Once a solution is obtained with a SAT solver \cite{audemard2009glucose, imms-sat18}, we can infer the colours and orientations for each particle species by listing variables  \(x_{s,p,c}^{C}\)  and  \(x_{s,p,o}^{O}\) that are true, and interaction matrix between colours is obtained by listing  \(x_{c_i,c_j}^{B}\) that are true. To explicitly prevent a particular undesired solution, which has been shown to form undesired structures, we can add a new additional clause that is a disjunction of  negation of variables   \(x_{c_i,c_j}^{B}\)  and  \(x_{s,p,c}^{C}\) that correspond to the undesired solution.

Another important difference to Refs.~\cite{romano2020designing,russo2021sat} is that the method presented here allows for bounded structures. This is done by adding species of type ``empty'' as a ``shell'' around the shape, as seen in Supp.~Figure~S11, to ensure that empty patches remain unbound.

\begin{table}[ht]
\centering
\caption{SAT variables and their descriptions}
\begin{tabular}{|l|p{6cm}|}
     \hline
     Variable & Description \\ [0.5ex] 
     \hline
     
    \(x_{l,p,o}^{A}\) & Patch \(p\) at position \(l\) has orientation \(o\) \\
    \(x_{c_i,c_j}^{B}\) &  Colour \(c_i\) is compatible with colour \(c_j\) \\
    \(x_{s,p,c}^{C}\) & Patch \(p\) on species \(s\) has colour \(c\) \\
    \(x_{p_1,o_1,p_2,o_2}^{D}\) & Patch \(p_1\), orientation \(o_1\) binds to patch \(p_2\), orientation \(o_2\) \\
    \(x_{l,p,c}^{F}\) & Patch \(p\) at position \(l\) has colour \(c\) \\
    \(x_{s,p,o}^{O}\) & Patch \(p\) on species \(s\) has orientation \(o\) \\
    \(x_{l,s,r}^{P}\) & Position \(l\) is occupied by species \(s\) rotated by \(r\) \\
     \hline
\end{tabular}
\end{table}

\begin{table}[ht]
\centering
\caption{SAT clauses definition}
\begin{tabular}{|l|l|l|}
    \hline
     Id & Clause & Boolean expression \\ [0.5ex] 
     \hline
     (i) & \(C^{B}_{c_i,c_j,c_k}\) & \(\neg x_{c_i,c_j}^{B} \lor \neg x_{c_i,c_k}^{B}\) \\ 
     (ii) &  \(C^{C}_{s,p,c_k,c_l}\) & \(\neg x_{s, p, c_k}^{C} \lor \neg x_{s, p, c_l}^{C}\) \\ 
     (iii) & \(C^{P}_{l, s_i, r_i, s_j, r_j}\)  & \(\neg x_{l,s_i,r_i}^{P} \lor \neg x_{l,s_j,r_j}^{P} \) \\ 
     (iv) & \(C^{BF}_{l_i,p_i,c_i,l_j,p_j,c_j}\) & \(\left(x_{l_i,p_i,c_i}^{F} \land x_{l_j,p_j,c_j}^{F} \right) \Rightarrow x_{c_i,c_j}^{B}\) \\ 
     (v) & \(C^{rotC}_{l,s,r,p,c}\) & \(x_{l,s,r}^{P} \Rightarrow \left(x_{l,p,c}^{F} \Leftrightarrow x_{s, \phi_r(p), c}^{C}\right)\) \\ 
     (vi) & \(C^{all s}_{s}\)  & \(\bigvee_{\forall l, r} x_{l,s,r}^{P}\) \\ 
     (vii) & \(C^{all c}_{c}\)  & \(\bigvee_{\forall s, p} x_{s,p,c}^{C}\) \\ 
     (iix) &  \(C^{O}_{s,p,o_k,o_l}\) & \(\neg x_{s, p, o_k}^{O} \lor \neg x_{s, p, o_l}^{O}\) \\ 
     (ix) & \(C^{DA}_{l_i,p_i,c_i,l_j,p_j,c_j}\) & \(\left(x_{l_i,p_i,c_i}^{A} \land x_{l_j,p_j,c_j}^{A} \right) \Rightarrow x_{p_i,c_i,p_j,c_j}^{D}\) \\ 
     (x) & \(C^{rotO}_{l,s,r,p,o}\) & \(x_{l,s,r}^{P} \Rightarrow \left(x_{l,p,o}^{A} \Leftrightarrow x_{s, \phi_r(p), o}^{O}\right)\) \\ 
     
     \hline
\end{tabular}
\end{table}


\subsection*{Substitution solving}
For solutions with $N_s$ close to the fully addressable solution, we also implement a substitution solving approach, which starts from fully addressable solution, identifies species with the same configuration of patches, and substitutes them with a single species. The substitution continues as long as the updated rule assembles correctly.


\subsection*{The patchy particle model}
\label{sec:patchyParticleModel}
In the molecular dynamics simulations, each particle is represented by a sphere covered by up to 6 patches at distance $R = 0.5$ distance units (d.u.) from the centre of the sphere. The positions of the patches, defined in terms of the orthonormal base associated with the patchy particle, are
$
 \mathbf{p}_1 = R \left( 0,1,0\right),\, \mathbf{p}_2 = R \left( 0,-1,0 \right),\,   
 \mathbf{p}_3 = R \left( 0,0,1 \right),\, \mathbf{p}_4 = R \left(  0, 0, -1 \right),\, 
 \mathbf{p}_5 = R \left( 1,0,0 \right),\, \mathbf{p}_6 = R \left( -1, 0, 0 \right),
$
corresponding to the faces of the cube. Each patch is additionally assigned an orientation $\mathbf{o}$, which is equal to
$\pm \mathbf{e}_1, \pm \mathbf{e}_2$, or $\pm \mathbf{e}_3$, which are the base vectors (or their negatives) of the orthonormal base of the particle. The orientation  $\mathbf{o}_a$ assigned to each patch $a$ satisfies $\mathbf{p}_a \cdot \mathbf{o}_a = 0$.

\begin{figure}[h]
\centering
\includegraphics[width=0.8\linewidth]{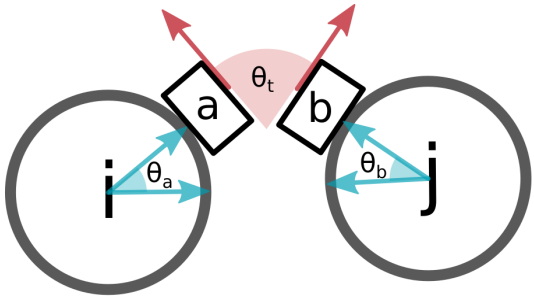}
\caption{Schematic of the patchy particle alignment angles. Note that, while the figure is drawn in 2D, the particles and vectors are in fact three-dimensional and are not restricted to the depicted plane. The angle \(\theta_a\) is measured between the vectors \(\mathbf{\hat{r}_{ij}}\) (pointing from particle \(i\) to particle \(j\)) and \(\mathbf{\hat{p}_a}\) (pointing from from particle \(i\) to its patch \(a\).  \(\theta_b\). Likewise, the angle \(\theta_b\) is measured between the vectors \(-\mathbf{\hat{r}_{ij}}\) (pointing from particle \(j\) to particle \(i\)) and \(\mathbf{\hat{p}_b}\) (pointing from from particle \(j\) to its patch \(n\). Finally the angle \(\theta_t\) is measured between the orientation vectors of the two patches, \(\mathbf{o}_{a}\) and \(\mathbf{o}_{b}\), which are always orthogonal to \(\mathbf{\hat{p}_a}\) and \(\mathbf{\hat{p}_b}\) respectively.
}
\label{fig:pp}
\end{figure}

\begin{figure}[h]
\includegraphics[width=\linewidth]{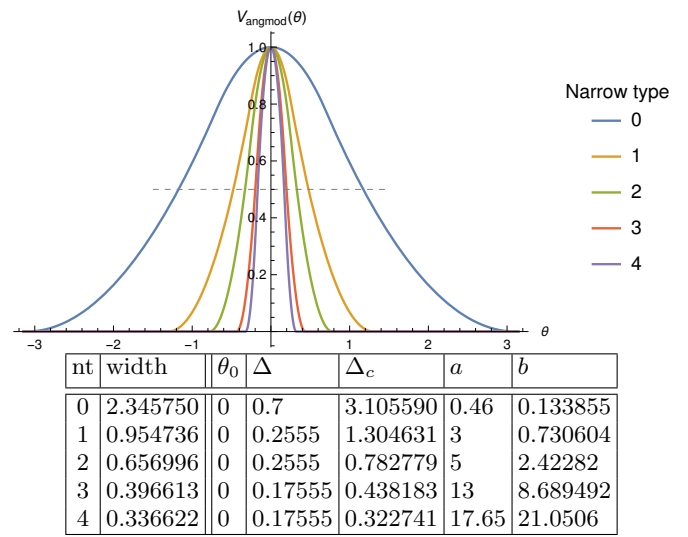}
\begin{tabular}{|c|l||l|l|l|l|l|}
    \hline
    nt & width & \(\theta_0\) & \(\Delta\) & \(\Delta_c\) & \(a\) & \(b\) \\ [0.5ex] 
    \hline\hline
    0 & 2.345750 & 0 & 0.7 & 3.105590 & 0.46 & 0.133855 \\
    1 & 0.954736 & 0 & 0.2555 & 1.304631 & 3 & 0.730604 \\
    2 & 0.656996 & 0 & 0.2555 & 0.782779 & 5 & 2.42282 \\
    3 & 0.396613 & 0 & 0.17555 & 0.438183 & 13 & 8.689492 \\
    4 & 0.336622 & 0 & 0.17555 & 0.322741 & 17.65 & 21.0506 \\
    \hline
\end{tabular}
\caption{Patchy particle narrow types, causing patches of different widths. The table shows the constants used for each narrow type, with the plot showing the different angular modulation potentials as a function of angle in radians. The width is measured between the two points intersecting the dashed line, where \(V_{\text{angmod}}=\frac{1}{2}\). The default narrow type \(0\) is the least narrow. }
\label{fig:narrow_types}
\end{figure}

We model the interaction between patches using point-like patch interaction, where the interaction between two patches $a$ and $b$ on two distinct particles $i$ and $j$ is given by the following interaction potential:
\begin{equation}
\label{eq_pointpatch}
V_{\rm patch}(\mathbf{r}_{ij},\mathbf{\Omega}_{ij}) = \delta_{ab} V_{\rm pdist}(r_p) V_{\rm angle}(\theta_a, \theta_b, \theta_t), 
\end{equation}
 where $\delta_{ab}$ is 1 if colours assigned to patch $a$ and $b$ can bind and 0 otherwise. The patch potential consists of two components: potential $V_{\rm pdist}$ that only depends on distance between the two patches $r_{ab}$, and $V_{\rm angle}(\theta_a, \theta_b, \theta_t)$ which depends on the mutual orientation $\mathbf{\Omega}_{ij}$ of the two particles as given by angles that are calculated as follows:
\begin{equation}
\cos \theta_a =  \mathbf{\hat{r}}_{ij} \cdot \mathbf{\hat{p}}_a, \quad \cos \theta_b =  \mathbf{-\hat{r}}_{ij} \cdot \mathbf{\hat{p}}_b, \quad 
\cos \theta_t =  \mathbf{o}_{a} \cdot \mathbf{o}_b,
\end{equation}
where the above vectors are normalised to 1, as indicated by the hat symbol. As seen in Figure~\ref{fig:pp}, angles $\theta_a$ and $\theta_b$ correspond to the angle between the normalised vector between the centres of mass of patchy particle $i$ and $j$ and the normalised vector pointing to patch $a$ or $b$ respectively. Angle $\theta_t$ corresponds to the angle between the orientations associated with the respective patches. The interaction potential is 
\begin{equation}
    V_{\rm angle}(\theta_a, \theta_b, \theta_t) = V_{\rm angmod}(\theta_a) V_{\rm angmod}(\theta_b)
    V_{\rm angmod}(\theta_t) ,
\end{equation}
where $V_{\rm angmod}$ is angular modulation function defined to be equal to one if the angle $\theta$ equals to the desired angle $\theta_0$ (which we set to $0$, requiring perfect alignment of the respective vectors), and parameters $a$ and  $\Delta$ then define the width of the potential $V_{\rm angmod}(\theta) = $ 
\begin{equation}
\begin{cases}
	V_{\rm mod}(\theta, a, \theta_0)  & \text{if $ \theta_0 - \Delta < \theta < \theta_0 + \Delta  $},\\
	V_{\rm smooth} (\theta, b, \theta_0 - \Delta_c) & \text{if $\theta_0 - \Delta_c < \theta < \theta_0-\Delta $},\\
	V_{\rm smooth} (\theta, b, \theta_0 + \Delta_c) & \text{if $\theta_0 + \Delta < \theta < \theta_0+\Delta_c$},\\
	0 & \text{otherwise}.
	\end{cases} 
\end{equation}
where the additional parameters $b$ and $\Delta_c$ are set so that the piece-wise function $V_{\rm angmod}(\theta)$ is differentiable (See Figure~\ref{fig:narrow_types}). The potentials used in the definition are 
\begin{equation}
\label{eq:smooth}
V_{\rm smooth} (x, b, x_c) = b(x_c - x)^2, 
\end{equation}
and
\begin{equation}
V_{\rm mod} (\theta, a, \theta_0) = 1 - a (\theta-\theta_0)^2.
\end{equation}

Unless otherwise specified, by default the widest patch (narrow type 0) was used, with 
$a = 0.46$ and $\Delta = 0.7$ for $V_{\rm angmod}$ as seen in Figure~\ref{fig:narrow_types}. For the simulations labeled as \textit{no torsion}, the interaction potential between patches as specified in Eq.\ref{eq_pointpatch} is replaced by interaction potential 
\begin{equation}
V_{\rm patch}^{\rm no\,torsion}(\mathbf{r}_{ij},\mathbf{\Omega}_{ij}) = \delta_{ab} V_{\rm pdist}(r_p),
\end{equation}
which is independent of angles that describe mutual patch orientation, and only depends on the distance between patches.

The additional distance-modulation term in the potential between a pair of patches on two distinct particles is 
\begin{equation}
    \label{eq_patch}
    V_{\rm pdist}(r_p) =  \begin{cases} 
                          -1.001  \exp{\left[- \left( \frac{r_p}{\alpha} \right)^{10}\right]} + C & \text{if $r_p \leq r_{\rm pmax}  $} \\
                          0 & \text{otherwise}
                        \end{cases}
\end{equation}
where $r_p$ is the distances between a pair of patches, and $\alpha = 0.12\, {\rm d.u.}$ sets the patch width. The constant $C$ is set so that $V_{\rm patch}(r_{\rm pmax}) = 0$ for $r_{\rm pmax} = 0.18\, \rm{d.u.}$. The patchy particles further interact through excluded volume interactions ensuring that two particles do not overlap:
\begin{equation}
    f_{\rm exc}(r,\epsilon,\sigma,r^{\star}) = \begin{cases}
	V_{\rm LJ}(r, \epsilon, \sigma) & \text{if $r < r^{\star} $},\\
	\epsilon V_{\rm smooth} (r, b, r^c) & \text{if $r^{\star} < r < r^c$},\\
	0 & \text{otherwise}.
	\end{cases} 
\end{equation}
where $r$ is the distance between the centers of mass of the patchy particles, and $\sigma$ is set to $2R = 1.0$ distance units, twice the desired radius of the patchy particle.
The repulsive potential is a piecewise function, consisting of the Lennard-Jones potential function:
\begin{equation} 
V_{\rm LJ}(r,  \sigma) = 8 \left[ \left(\frac{\sigma}{r}\right) ^{12} - \left(\frac{\sigma}{r}\right) ^{6} \right].
\end{equation}
that is truncated using a quadratic smoothing function from \eqref{eq:smooth}, with $b$ and $x_c$ set so that the potential is a differentiable function that is equal to $0$ after a specified cutoff distance $r^c = 0.8$.  

The patchy particle system was simulated using rigid-body Molecular Dynamics with an Andersen-like thermostat~\cite{russo2009reversible}. During the simulation, each patch was only able to be bound to one other patch at the time, and if the binding energy between a pair of patches, as given by Eq.~\eqref{eq_patch}, is smaller than 0, none of the patches can bind to any other patch until their pair interaction potential is again 0.
The modified oxDNA implementation used for the patchy particle simulations can be found at \href{https://github.com/Akodiat/oxDNA_torsion}{https://github.com/Akodiat/oxDNA\_torsion}.

\subsection*{Yield calculation}
The patchy particle simulation yields are calculated using edge-induced subgraph isomorphism \cite{networkx}. 
We annotate \(\sigma(G_a,G_b) == \text{True}\) if the graph \(G_a\) is an edge-induced subgraph of the graph \(G_b\).
The connectivity graph \(G_i\) for each assembled particle cluster is compared to the graph for the intended shape \(G_{\rm correct}\). If \(G_i\) is a large enough edge-induced subgraph of \(G_{\text{correct}}\), meaning that its particles are connected like a large enough subset (the current results use a 75\% size cutoff) of the correct shape, it contributes to the yield with its fraction of correctly assembled particles:
\begin{equation}
    Y_{c} = \sum_{G_i \in c} \begin{cases} 
           \frac{\left|N(G_i)\right|}{\left|N(G_{\text{correct}})\right|} & \text{if } \sigma(G_i,G_{\text{correct}}) \\
                          0 & \text{otherwise}
                        \end{cases}
\end{equation}

The subgraph calculation becomes very compute-intensive for larger and more highly connected graphs, so for the \(4\times4\times4\) solid cube we instead check the cluster's species composition, making sure that its species are a subset (with duplicates) of the intended shape. For the minimal solution, this underestimates the yield for some temperatures compared to the subgraph measure, as the solution can produce correct shapes with varying species compositions (but some mismatched connections). 

The other exception is used in the case of the werewolf design, where the connectivity graph of the human structure is a subgraph of the connectivity graph of the wolf structure. To tell the two shapes apart, we therefore check both the connectivity graph and the species composition.

\subsection*{DNA nanostructure design and simulation}
The individual DNA cube origami was exported from the ATHENA tool \cite{jun2021rapid} to oxDNA and relaxed according to the protocol in Ref.~\cite{Bohlin2022}. Using the template provided from the stochastic polycube simulation, we used an export script to position the origami cubes at the assembled orientations and positions, ligating the ssDNA overhang positions with connecting helices at every connected patch. We then relaxed the structure again, and ran a production run with the oxDNA model at temperature $20^\circ$C for \(10^9\) simulation steps, which corresponds approximately to \(10^{-3}\) seconds in experiment \cite{vsulc2012sequence,Snodin2015,ouldridge2011structural}.
To obtain a mean structure of the production run, every configuration in the production trajectory was aligned to a randomly chosen configuration from the production run. Next we compute an average of all coordinates, and calculated deviations from them over aligned structures from the production run to obtain the flexibility profile, using the tools from Ref.~\cite{poppleton2020design}.


\section*{Acknowledgements}
We acknowledge the support from ONR grant no N000142012094. This project has received funding from the European Union’s Horizon 2020 research and innovation programme under the Marie Skłodowska-Curie grant agreement No 765703. We further acknowledge Research Computing at Arizona State University for providing HPC resources that have contributed to the research results reported within this paper. We thank Michael Matthies for his help with the design and assembly of DNA nanostructures and to Flavio Romano, Lukáš Kroc, Lorenzo Rovigatti and John Russo for helpful discussions and their contributions of SAT-assembly and oxDNA code development.


\newpage
\bibliography{pnas-sample}

\end{document}